# COMMISSIONING OF THE 1.4 MeV/u HIGH CURRENT HEAVY ION LINAC AT GSI

W. Barth, GSI, Planckstr.1, 64291 Darmstadt, Germany

*Abstract*

The disassembly of the Unilac prestripper linac of the Wideröe type took place at the beginning of 1999. An increase of more than two orders of magnitude in particle number for the most heavy elements in the SIS had to be gained. Since that time the new High Current Injector (HSI) consisting of H-type RFQ and DTL-structures for dual beam operation was installed and successfully commissioned. The High Charge Injector (HLI) supplied the main linac during that time. Simultaneously conditioning and running in of the rf-transmitters and rf-structures were done. The HSI commissioning strategy included beam investigation after each transport and acceleration section, using a versatile diagnostic test stand. Results of the extensive commissioning measurements (e.g. transverse emittance, bunch width, beam transmission) behind LEBT, RFQ, Super Lens, IH tank I and II and stripping section will be discussed. An $^{40}Ar^{1+}$ beam coming from a MUCIS ion source was used to fill the linac up to the theoretical space charge limit. Routine operation started in November 1999.

## 1 INTRODUCTION

Table 1: Specified beam parameters at Unilac and SIS injection, exemplary for a uranium beam.

|  | HSI entrance | HSI exit | Alvarez entrance | SIS injection |
|---|---|---|---|---|
| Ion species | $^{238}U^{4+}$ | $^{238}U^{4+}$ | $^{238}U^{28+}$ | $^{238}U^{73+}$ |
| El. Current [mA] | 16.5 | 15 | 12.5 | 4.6 |
| Part. per 100µs pulse | $2.6 \cdot 10^{12}$ | $2.3 \cdot 10^{12}$ | $2.8 \cdot 10^{11}$ | $4.2 \cdot 10^{10}$ |
| Energy [MeV/u] | 0.0022 | 1.4 | 1.4 | 11.4 |
| $\Delta W/W$ | - | $\pm 4 \cdot 10^{-3}$ | $\pm 2 \cdot 10^{-3}$ | $\pm 2 \cdot 10^{-3}$ |
| $\varepsilon_{n,x}$ [mm mrad] | 0.3 | 0.5 | 0.75 | 0.8 |
| $\varepsilon_{n,y}$ [mm mrad] | 0.3 | 0.5 | 0.75 | 2.5 |

The original Unilac was not dedicated as a synchrotron injector, fulfilling all requirements due to high intensities (especially for mass number above 150). In 1994 the vision of a new High Current Injector (HSI) was drafted. This injector should provide an increase of beam intensities by 2.5 orders of magnitude filling the synchrotron up to its space charge limit for all ions – including uranium [1]. An increase of the accelerating gain by a factor of 2.5 is necessary to accelerate ion species up to maximum A/q-values of 65 ($^{130}Xe^{2+}$) within the given length of the former Wideröe injector. For a 15mA $^{238}U^{4+}$ beam out of the HSI $4 \cdot 10^{10}$ $U^{73+}$ particles should be delivered to the SIS during 100µs. This means that the SIS space charge is reached by a 20 turn injection into the horizontal phase space. The required parameters of beam quality are summarised in Tab. 1 for the uranium case.

## 2 HSI LAYOUT

The beam of the new High Current Injector is stripped and injected into the Alvarez accelerator, which is approx. 30 years in operation now. It was predicted by PARMILA-calculation and confirmed by beam tests that the Alvarez accelerates highly space charge dominated ion beams coming from the new HSI without any significant particle loss and without decrease in brilliance [2].

The new injector is illustrated in Fig. 1, a more detailed description of the main acceleration and transport-sections is given in the following.

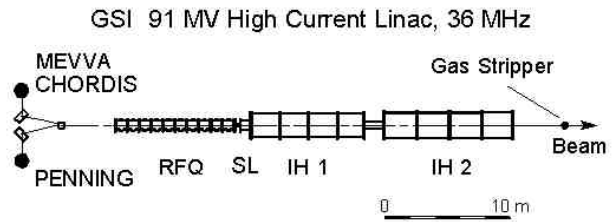

Fig. 1: The HSI as a scheme.

### 2.1 Ion Source and LEBT section

The HSI is fed by two ion source terminals: One of them is yet since many years housing a Penning ion source (PIG), generating intermediate charge state ion beams with a duty cycle up to 30% (50Hz, 6ms), limited to an A/q value of 24. The other one is upgraded as a high current terminal; optionally the MUCIS source or the MEVVA source is installable, providing short intense macro pulses ($\leq$ 1.2ms, $\leq$ 16Hz) as required for the linac operation with ion beams of high magnetic rigidity (A/q$\leq$65) [3]. The existing LEBT [4] section down to the switching magnet was surveyed as basically useful for loss-free transport of high intense beams, assuming a high degree of space charge compensation. The section from the switching magnet to the RFQ entrance was rebuilt as a "50Hz-pulseable" beamline, allowing for a two beam operation now. The RFQ-matching condition of a double-waist at a very small transverse beam diameter of 5mm is accomplished by a quadrupole quartet. A beam chopper with a rise time less than 500ns is located close to the entrance of the RFQ.

## 2.2 36 MHz IH-RFQ and Matching to IH-DTL

The 9.35m long 36 MHz IH-RFQ [5] accelerates the ion beam from 2.2 keV/u up to 120 keV/u, where a voltage amplitude of 137 kV and a max. surface field of 28 MV/m is necessary to provide the required accelerating gain. The postulated mechanical precision of electrodes [9] should be better than 0.05 mm along each module, obtaining a high transmission rate. The time for rf-conditioning up to now was less than 500 h. The reduction of a primarily large dark current contribution was observed in so much that 90% of the design field are reached now routinely to obtain optimum operation for $U^{4+}$ beams. The matching to the IH-DTL is done with a very short (0.8m) 11 cell adapter RFQ [6], with large aperture and a synchronous phase of $-90^0$ (Super Lens). The surface field is 26 MV/m at a design vane voltage of 212 kV.

## 2.3 83 MV IH-DTL

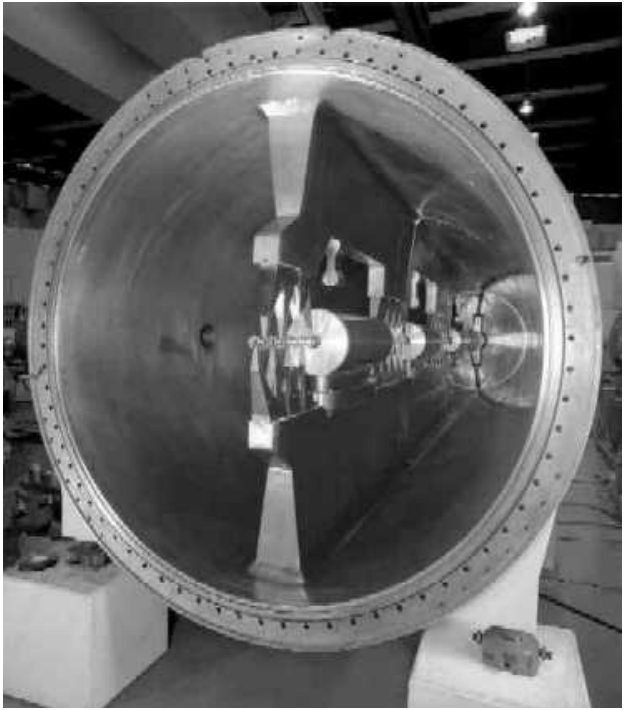

Fig. 2: A view into IH1: Four KONUS sections, connected by the three large drift tubes each housing a quadrupole triplet, can be seen.

The IH-DTL [7] consists of two separate tanks (9.1m and 10.3m long), connected by an intertank section. The final beam energy of IH1 is 0.743 MeV/u, while IH2 accelerates the ion beam to the full HSI-energy of 1.4 MeV/u. The whole DTL is structured into six KONUS ("Kombinierte Nullgrad Struktur") sections, mainly consisting of a rebuncher section operating at a synchronous phase of $-35^0$, followed by the acceleration stage ($\Phi s=0^0$), where the synchronous particle is injected with a surplus in energy resulting in a high accelerating rate. Finally in each section a quadrupole triplet provide the transverse focusing. Tank 1 houses four accelerating sections with three internal triplet lenses as shown in Fig.2. Behind the external triplet tank 2 contains two accelerating sections and one external triplet between. The rf-conditioning of the structures results in lower dark current contributions regarding rf power losses when compared to the more critical RFQ.

## 2.4 1.4 MeV/u and 11.4 MeV/u stripper section

Two quadrupole doublets match the 1.4 MeV/u HSI-beam to the gas stripper [8]. No additionally focusing device is necessary to feed the new charge state separator system. If as a worst case a 15mA $U^{4+}$ ion beam is stripped, $U^{28+}$ should purely separated from the neighbouring charge states under extremely high space charge conditions (105 emA total pulse current). After the rebuilt of the stripper region a multi-pulse mode [9] from the different injectors is possible. In the 11.4 MeV/u stripper region [10] of the transfer line to the synchrotron the beam power in case of an intense uranium beam is approximately as high as in the gas stripper section. In order to cope with the potentially disastrous beam load on the stripper foil a magnetic sweeper system had been established, saving the foil, while a significant decrease of beam quality due to stripping effects should not occur. A new short charge state separator system, to be installed directly behind the foil stripper, is still under construction.

## 3 RF SYSTEM

Supplying the 36 MHz structures, rf amplifiers with a peak power of about 2 MW had to be installed, while the 27 MHz rf equipment became dispensable. Additionally redesigned fast amplitude- and phase controls were built along the whole Unilac, enabling the operation with high beam loading. The five 200 kW amplifiers were externally built (TOMCAST AG, Switzerland) and partly utilised as pre amplifiers for the 2 MW end stages and as rf provider for the Super Lens and a rebuncher. The in house developed end stages are powered by the Siemens tetrode RS2074 SK; they feed the RFQ and the two IH cavities. The assembly of all rf-components took place just in time and the whole system shows excellent reliability [11], [12].

## 4 BEAM DIAGNOSTIC TEST BENCH

A mobile test bench [13] was designed and already used for the stepwise performed beam commissioning of LEBT, RFQ, Super Lens and IH-DTL. It was equipped with in part newly designed beam diagnostics, as shown in Fig.3: four segmented capacitive pick-up probes, beam transformers, a profile grid in combination with a non-destructive residual gas ionisation profile monitor, a slit-grid emittance measurement device, as well as a first time used pepperpot system and a particle detector for bunch structure observation. At the end of the test bench the beam was dumped in a cooled Faraday cup.

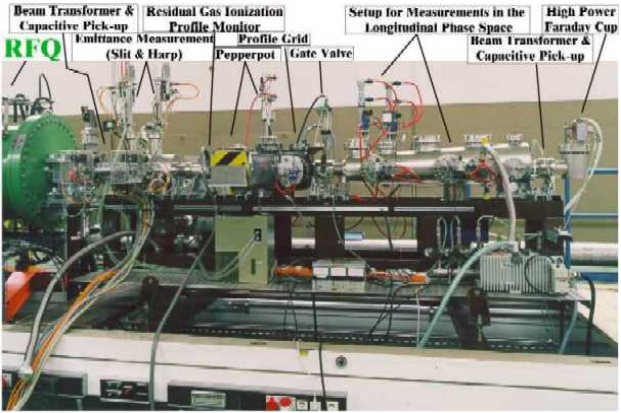

Fig. 3: Beam diagnostic test bench during beam commissioning of the 36 MHz IH-RFQ. [13]

## 5 TIME SCHEDULE&ACHIEVEMENT

Table 2.: HSI-Assembly&Commissioning milestones

| Dec. 98 | Last operation-shift with Wideröe injector |
|---|---|
| Jan.-Feb. 99 | Disassembly of Wideröe and rf, installation of LEBT section |
| March 99 | Successful commissioning of LEBT |
| April-May 99 | Mounting IH-RFQ and first acceleration up to 120 keV/u |
| June 99 | Beam tests with Super Lens, achieving 10 mA $Ar^{1+}$ at RFQ exit |
| July 99 | Assembly of IH1, verification of beam acceleration up to 743 keV/u |
| August 99 | Completing HSI with IH2 and stripper section |
| 2.Sept. 99 | Proof of acceleration up to 1.4 MeV/u, further on: 80% IH-transmission for highest argon intensities (8 mA) |
| October 99 | Upgrade of transfer line to SIS and mounting of matching section to Alvarez |
| November 99 | Establishing three beam operation, complete Alvarez transmission at highest current |
| Since Nov. 99 | HSI in routine operation |
| February 2000 | Achievement of the 90%-rf levels, first 1.4 MeV/u $U^{4+}$ beam (3 mA) |

The mounting and commissioning of the new injector took place in the first 9 months of 1999 – the assembly was done in 6 steps, each subsequently completed with a "two week beam-commissioning period". Regular beam operation started end of November, just in time as scheduled a long time in advance; the milestones of mounting and commissioning are summarised in table 2.

## 6 BEAM COMMISSIONING RESULTS

### 6.1 TOF-Measurements

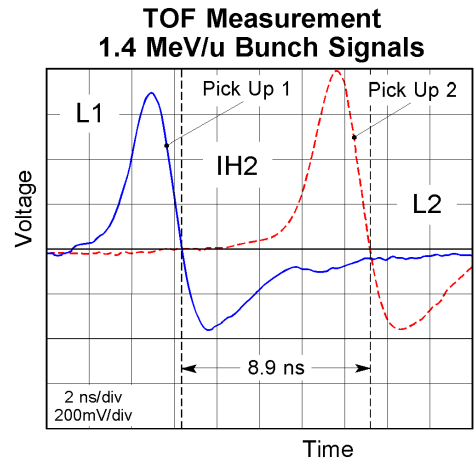

Fig. 4: Phase probe signals behind IH2 structure.

After each step the correct beam energy was verified by a time of flight measurement using the signals of two pick-ups in a well-known distance. As an example Fig. 4 shows an oscilloscope view of two signals, received by phase probes, placed after the IH2 structure (1.396 MeV/u). The achievable accuracy $\Delta W/W$ is less than +/-0.12%. The evaluation of difference signals allows an on-line monitoring of the beam position for a higher intensity.

### 6.2 Phase probe signals

Considering a max. beam power of 1.3 MW inside a macro-pulse at the HSI-energy of 1.4 MeV/u - leading to melting and evaporation of hit material after some µs - it is necessary to limit the rise time of the macro pulse to a value as short as possible [9]. As shown in Fig.5 the measured rise time of the phase probe signal after the RFQ is of the same amount as the design value of the chopper (0.5 μs). No significant increase of rise time takes place along the whole HSI, demonstrating good working rf control loops with respect to beam loading of the cavities. Macro-pulse shape investigations were carried out for $Ar^{1+}$ close to the current limit, as well as for many different other ion species and intensities (for example 6.5 mA $^{238}U^{4+}$ from the MEVVA source [15]) and showed no considerable difference.

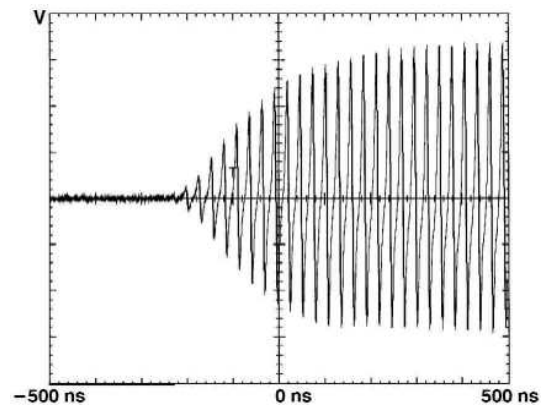

Fig. 5: Rise time of the phase probe signal after RFQ [14]

## 6.3 Particle loss and transversal emittance

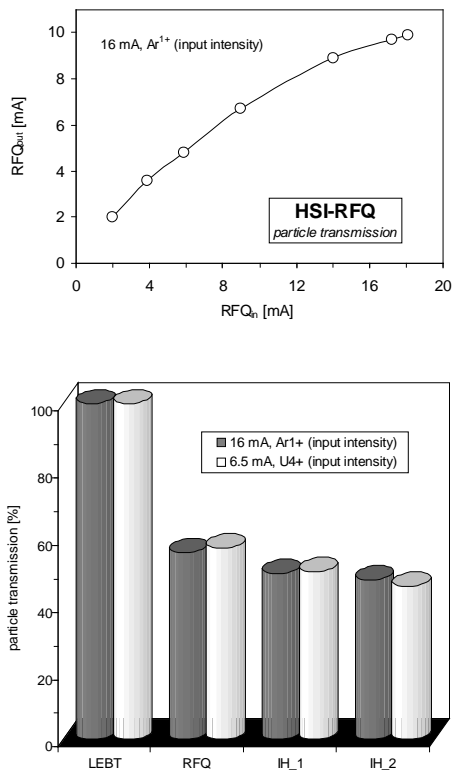

Fig. 6: Output current of the RFQ as a function of the input intensity (top) and particle transmission along the HSI (bottom) for $Ar^{1+}$ (MUCIS) and $U^{4+}$ (MEVVA)

Fig. 6 (above) represents the measurement of the $Ar^{1+}$ current by a beam transformer behind of the RFQ as a function of the injected beam in front of the quadruplet lense. The RFQ-matching was optimised for the high intensity case (18 mA) in such a way that the design current limit of 10 mA was reached [16]. The controlled intensity reduction was done by cutting the horizontal phase space distribution with slits and without any tuning of the RFQ matching, resulting in full particle transmission from 4 mA downwards. The transmission decrease to roughly 55 % at the theoretical current limit (10 mA) can not be completely explained by high transversal input emittance: The emittance measurements resulted in normalised 90%-emittances from 0.25 π·mm·mrad up to 0.45 mm·mrad without any significant influence to RFQ-transmission. Mismatch problems due to space charge effects or misalignment inside the RFQ are not ruled out to lead to particle loss. The RFQ is the bottleneck as figured out and emerged by many measurements; the transmission of the IH-DTL is better than 90% over a wide range of beam intensities and ion species (Fig. 6, beneath). For the $U^{4+}$ case 3 mA after the HSI were reached, while the particle transmission is close to the amount of the space charge dominated argon beam. The high intensity-fluctuation due to the behaviour of the MEVVA source is about ±25 % (30 following macropulses), without a significance influence to the beam emittance. The pulse-reproducibility of position and amount of the transverse emittance at the HSI exit is better than 4 % (verified by pepperpot emittance measurements). The beam loading for instance for RFQ (44 kW) and IH2 (147 kW) measured by the additional rf-power needed for the acceleration is close to the theoretical values.

## 6.4 Emittance growth

Measurements of the transversal emittance for the several energy steps of the HSI (and at 11.4 MeV/u) were done exclusively with a slit-grid device for short pulses. The input emittance was measured before the quadrupole quartet. For 120 keV/u, 750 keV/u and 1.4 MeV/u the beam was transported to a measurement device in the gas stripper region [17]; another device is placed after the Alvarez. Fig. 7 summarises the measured emittance data for an $Ar^{1+}$ beam with 10 mA at RFQ injection and 6.5 mA at the HSI exit. The $Ar^{10+}$ current (after stripping and charge state analysis) came up to 7 mA by gas stripper density variation. The measurements agreed to the calculation [18], if a measuring error of about ±15 % is taken into account.

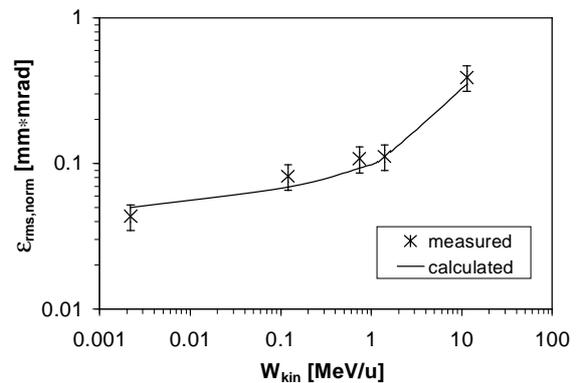

Fig. 7 Measurement of the horizontal emittance along the Unilac

## 6.5 Bunch shape

Bunch shape measurements were done by the use of diamond detectors, whereas the ion beam (here $Ar^{1+}$)

passes a thin Au-foil – the "Rutherford"-scattered particles hit the detector below a small angle. The bunch shape is obtained by measuring the arrival time of the particles against a reference [19]. It was even possible to observe the typical "zero current" phase space distribution in longitudinal plane, leading to intensity peaks in the center and at the beginning (resp. at the end) of the measured bunch shape. This effect is still present after accelerating the beam up to 0.743 MeV/u, and after transport to the stripper region (Fig. 8). Regardless the higher defocusing due to space charge forces in the high current case the bunch is shorter and without the significant "low current"-structure. At full HSI energy the beam is very well bunched. Independently from space charge effects along the HSI, the bunch is small enough to be matched to the poststripper by the two rebunchers operated at 36 MHz and 108 MHz, respectively.

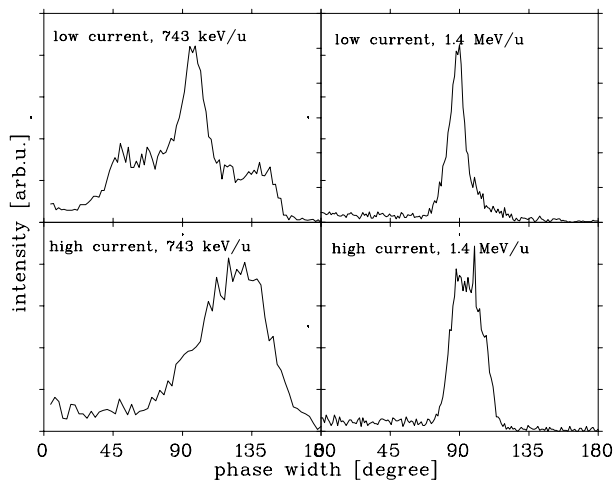

Fig. 8: Bunch shape measurements for different beam energies and intensities

## 7 CONCLUSION

The new High Current Injector was mounted and commissioned with great success. The measured beam parameters, as energy, bunch width, energy spread, transverse emittance fit very well to the calculation. The transmission at beam currents up to 40 % of the design intensity is close to 100%. Significant particle losses due to insufficient understanding problems are observable at the space charge limit. Within the intrinsic error bars of measurement the emittance growth in the high current case is as predicted by simulation. So far the rf levels are high enough to provide a stable operation with $U^{4+}$. First beam experiments with medium uranium intensity, feeding the HSI with the MEVVA beam, showed no significant deterioration of beam quality. Since November 1999 the HSI delivers beam to experiments in routine operation.

## 8 ACKNOWLEDGEMENTS